# Energy Options for Future Humans on Titan

Amanda R Hendrix[1*] and Yuk L Yung[2]

[1]Planetary Science Institute, Tucson, Arizona, USA

[2]Division of Planetary and Geological Sciences, Caltech, Pasadena, California, USA

[*]**Corresponding author:** Hendrix A, Planetary Science Institute, Tucson, Arizona, USA, Tel: 323-800-2700; E-mail: arh@psi.edu

## Abstract

We review the possibilities for *in situ* energy resources on Titan for use by future humans, including chemical, nuclear, wind, solar, geothermal and hydropower. All of these options, with the possible exception of geothermal, represent effective sources of power. Combustion of methane (after electrolysis of the native water), in combination with another source of power such as nuclear, is a viable option; another chemical source of energy is the hydrogenation of acetylene. The large seas Kraken and Ligeia potentially represent effective sources of hydropower. Wind power, particularly at altitudes ~40 km, is expected to be productive. Despite the distance from the sun and the absorbing atmosphere, solar power is (as on Earth) an extremely efficient source of power on Titan.

## Introduction

Once propulsion challenges are overcome, allowing humans to travel great distances quickly without incurring significant radiation damage, Saturn's moon Titan is the optimal location in the solar system for an off-Earth human settlement. It has Earth-like qualities and a thick atmosphere that provides shielding from damaging radiation [1,2] unlike any other solid surface location in the solar system. Given the distance from Earth (~$1.3 \times 10^9$ km), such a settlement must be self-sustainable, and in particular, humans will need to produce oxygen to breathe and provide heating for habitats. The abundance of hydrocarbons and availability of wind and hydro-based power give this self-sustainability a high likelihood. Here we discuss such energy options, from *In Situ* Resource Utilization (ISRU). This paper represents an exercise to provide a first look at energy options without delving into tedious detail on each one (at a time when questions about specific Titan conditions still exist), to provide readers with estimates of energy options for future study. Clearly, many details will need to be worked out in the coming decades, such as equipment mass required for delivery to Titan for any one of these systems.

## Nuclear

A nuclear power source brought to Titan from Earth would last several decades; for instance, $^{238}$Pu decays with a half-life of 88 years. After that, radioactive material can be extracted from silicate rocks; about 50% of Titan's mass is silicates [3]. Radiogenic argon (Ar) in Titan's atmosphere, likely a product of decay of $^{40}$K in the interior, has been observed by the Huygens GCMS [4] and the Cassini Orbiter Ion and Neutral Mass Spectrometer [5]. Such materials are likely to be deep in the interior assuming substantial differentiation, under hundreds of kilometers of ice and water, and thus extraction of radioactive materials would require significant effort and energy which we do not attempt to estimate here but note as a possibility.

## Chemical

Species in Titan's atmosphere and at the surface and subsurface can be utilized to produce chemical energy. Nitrogen makes up ~95% of the atmosphere. The abundance of methane ($CH_4$) increases with decreasing altitude below the tropopause at 32 km, leveling off at ~4.9% between 8 km and the surface [4]. This small amount of methane is responsible for keeping the nitrogen in the gas phase; without the methane, much of the nitrogen in the stratosphere and troposphere would condense out [6]. Trace amounts of other hydrocarbons, e.g. ethane ($C_2H_6$), diacetylene ($C_4H_2$), methylacetylene ($CH_3C_2H$), acetylene ($C_2H_2$) and propane ($C_3H_8$), and of other gases, such as cyanoacetylene ($HC_3N$), hydrogen cyanide (HCN), carbon dioxide ($CO_2$), carbon monoxide (CO), cyanogen ($C_2N_2$), argon and helium are also present [4,7]; $H_2$ is also present at larger abundances [8]. Complex $N_2/CH_4$ chemistry involving photons and energetic electrons in the atmosphere leads to the production of organic haze particles, which act as condensation nuclei and eventually settle to the surface as sediments ("tholins"). Titan's dune fields [9] are likely composed of organics that originated in the atmosphere. Methane/ethane rainfall occurs periodically on Titan [10], and liquid hydrocarbons fill the polar lakes and seas [11,12]. Water ice is available in abundance and makes up the bulk of Titan's interior [3]. All of these species represent potential resources.

Several options are available for producing energy via chemical pathways on Titan. Combustion of hydrocarbons alone (e.g. reactions 3, 4 in Table 1) is not efficient on Titan because it costs more energy to do the electrolysis needed to get the $O_2$ (reaction 5). In practice, a nuclear source could be used to run an electrolysis plant while chemical energy is produced via combustion separately. Alternatively, hydrogenation of acetylene (reaction 1) [13] is a viable exothermic option at Titan conditions, producing 376 kJ/mol of energy. Both acetylene and $H_2$ could be extracted from the atmosphere; alternately, pyrolysis of $CH_4$ (reaction 8) could be performed to produce $C_2H_2$ (though it would cost energy to do the pyrolysis). [On Mars [14], sources of $H_2$ include ferrous-ion reduction of $H_2O$ to $H_2$ during serpentinization, and photochemical dissociation of $H_2O$ in the atmosphere.] We note that an alternative (or additional option) for producing $O_2$ for combustion/respiration would be to use a greenhouse full of plants/algae to produce $O_2$; this would, however, require energy to operate and heat the greenhouse. We anticipate that, given the low efficiency of photosynthesis, genetically engineered plants/algae will be an attractive option for future Titan residents, but that is the subject of another paper.

We also consider possible uses of the abundant atmospheric nitrogen. For instance, hydrogenation of nitrogen (Table 1 reaction 10) is exothermic, producing ~92 kJ/mol of energy.

| Reaction number | Reaction | Energy |
|---|---|---|
| 1 | $C_2H_2 + 3H_2 \rightarrow 2CH_4$ (hydrogenation of acetylene) | -376.3 kJ/mol (exothermic) |
| 2 | $CO_2 + 4H_2 \rightarrow CH_4 + 2H_2O(g)$ (Sabatier process; methanogenesis) | -164.9 kJ/mol (exothermic) |
| 3 | $CH_4 + 2O_2 \rightarrow CO_2 + 2H_2O(g)$ (combustion of methane) | -802.3 kJ/mol |
| 4 | $2C_2H_6 + 7O_2 \rightarrow 4CO_2 + 6H_2O$ (combustion of ethane) | -2857.2 kJ/mol |
| 5 | $4H_2O \rightarrow 4H_2 + 2O_2$ electrolysis of $H_2O$ | 967.2 kJ/mol (241.8 x 4) |
| 6 | $2CO_2 \rightarrow 2CO + O_2$ (3000°C) | (endothermic) |
| 7 | $2CH_4 \rightarrow C_2H_2 + H_2$ (pyrolysis ~1000°C) | (endothermic) |
| 8 | $H_2O$ ice sublimation | 54 kJ/mol |
| 9 | $H_2O$ ice melting (heat of fusion) $H_2O$ (s) → $H_2O$ (l) | 6.01 kJ/mol |
| 10 | $N_2 + 3H_2 \rightarrow 2NH_3$ | -92.4 kJ/mol (exothermic) |

**Table 1:** Chemical energy reactions.

## Hydropower

Titan's abundant lakes and seas of methane and ethane can be used for hydropower, by creating a system to make the fluid run downhill. This is essentially a self-sustaining process, as the lakes are expected to be recharged when it rains and when the obliquity changes [15]. Based on the topographic information currently available, construction of such a system may be a significant engineering feat, given that the polar regions, home to Titan's lakes and seas, are topographically lower than lower latitudes [16]. This may not rule out the feasibility of a hydropower system, but higher-resolution topographic data are needed. Here we outline the possibilities for such a system.

The amount of power generated by a system is given by:

$P = \eta \rho Qgh$

Where $\eta$ is the efficiency of the turbine, $\rho$ is the density of the fluid, Q is the flow rate, g is the gravitational acceleration and h is the height difference. As an example, we use $\eta=0.85$, $\rho=660$ kg/m$^3$ [17], and h=145 m. For the volumetric flow rate Q, the standard value used for water on Earth is 80 m$^3$/s; given the estimated viscosity of the fluids in Titan's lakes (Hayes et al. [18] give a range of viscosities on Titan of 0.003 to 0.03 cm$^2$/s, compared with Earth's $H_2O$ viscosity range of 0.0084-0.0184 cm$^2$/s), we estimate a range in flow rates on Titan of 40 to 160 m$^3$/s. The gravitational acceleration at Titan is 14% of Earth's gravity, or 1.37 m/s$^2$. This Titan methanopower system would produce ~9 MW of power (Table 2). In comparison, for a similar system on Earth running on water, the power production would be 97 MW.

Estimating the depth of Titan's largest sea, Kraken Mare, at 500 to 1000 m [19], its length at 1100 km and an average width of 500 km, the volume of Kraken is estimated at $2.75-5.5 \times 10^{14}$ m$^3$. This results in some $3.1-6.1 \times 10^{19}$ J of energy produced by dropping the entire contents of the sea over a height of 145 m. Such a system could provide power (constrained by the volume of Kraken Mare) for 53,900 to 431,250 Earth years (1840 to 14,740 Titan years). In comparison, Earth's Lake Superior, which has a volume smaller than that of Kraken ($8.7 \times 10^{12}$ m$^3$), would produce $\sim 1 \times 10^{19}$ J of energy for a similar system for a duration of 3450 years. On Titan, if higher resolution topographic data indicate that there are no locations where a natural drop occurs and liquid flows downhill out of a lake, then such a system may require major engineering work which is out of the scope of this report to detail fully. We estimate that it would require cutting into the edge of the lake/sea to create the drop and then cutting into the landscape away from the lake/sea to allow the liquid to flow away, in a canal or pipeline. Without better knowledge of Titan's topography, we do not attempt in this report to estimate a precise location for such a project. More information is needed on Titan's topographic surface structure to plan out where such a system might be installed in the future.

| Parameter | Titan | Earth |
|---|---|---|
| Viscosity vf | 0.003-0.03 cm$^2$/s [16] | 0.0084-0.0184 cm$^2$/s [16] |
| Density $\rho$ | 660 kg/m$^3$ [18] | 1000 kg/m$^3$ |
| Gravity g | 1.37 m/s$^2$ | 9.81 m/s$^2$ |
| Flow rate Q | 40-160 m$^3$/s* | 80 m$^3$/s |
| Power** | 4.5-18 MW | 97 MW |
| Duration | 53,900-431,250 years (1840-14,740 Titan years) | 3450 years |
| Volume | $2.75-5.5 \times 10^{14}$ m$^3$ (Kraken Mare) | $8.7 \times 10^{12}$ m$^3$ (Lake Superior) |
| Total energy | $3.1-6.1 \times 10^{19}$ J | $1 \times 10^{19}$ J |

*Estimate based on range in viscosities and compared with Earth's standard flow rate
**Assuming $\eta=0.85$ and h=145 m

**Table 2:** Hydropower.

## Wind

Wind speeds at Titan's surface are minimal. Cassini instruments have mostly observed a lack of wind waves on the lakes and seas [18], though roughness on the sea Punga Mare has been interpreted as potentially due to winds of ~0.76 m/s [20]. However, dunes are present at mid latitudes, indicating the presence of some level of wind, at least in a recent epoch. Currently, surface wind speeds are estimated to be 0.5-1 m/s [21].

Power generated by wind turbines is given by:

$$P = 0.5 \times \eta \times A \times \rho \times v^3$$

Where $\eta$ is the efficiency (which accounts for intermittent wind), A is the area of the blade, v is the wind speed and $\rho$ is the air density. On Titan, the air density is roughly 5X that on Earth, but wind speeds are significantly lower. We use $\rho$=5 kg/m$^3$ [18], $\eta$=0.2 and v=0.5-1 m/s. Assuming turbine rotor diameters of 40-90 m, we estimate wind power production in the 79 W-3.2 kW range. (On Earth, with ~20 m/s wind speeds, the same sized wind machines would produce 1-5 MW of power.)

Wind speeds are higher on Titan (~2 m/s) at ~3 km altitude [21] and even higher (~20 m/s) at 40 km altitude [22]; tethered balloon or blimp-borne [23,24] power-generating windmills could be feasible to access these higher wind speeds for greater energy generation, on the order of hundreds of MW.

### Solar

Sunlight is a source of free and sustainable energy, though the solar flux at Titan's surface is limited. The amount of sunlight reaching Titan is ~1/81-1/100 that of the solar flux reaching Earth, due to Titan's distance from the Sun of 9-10 AU. We estimate the amount of solar energy available at Titan's surface by scaling down from Earth. At the top of Earth's atmosphere, the average solar energy is 1400 J/m$^2$-s. At the top of Titan's atmosphere, this scales to 14-17 J/m$^2$-s. Titan's atmospheric transmission depends on wavelength: the red and near-infrared are transmitted (minus methane absorption) whereas blue light is absorbed [25]; here we assume that 10% of the solar flux makes it to Titan's surface. In reviewing the response function of various photovoltaic materials, we estimate that Titan's transmitted spectrum is best matched by the response of amorphous silicon or perhaps cadmium telluride (CdTe) photovoltaic material. The efficiency of these material is in the ~13-20% range but the performance at Titan temperatures is unknown. To be conservative in this initial, simplified exercise, we estimate the efficiency at 10%. We also consider that for any low-mid latitude location on the surface, the sun is up for ~1/3 of the day. (This does not consider seasonal variations or eclipses by Saturn.)

If we consider a future Titan settlement with roughly the population of the US (which has a surface area ~10.8% of Titan's), we can assume that they might consume roughly the amount of energy currently consumed in the US, ~1.4 × 10$^{19}$ J/year, or 4.2 × 10$^{20}$ J/Titan year. To meet this need at Titan, with ~1.4-1.7 J/m$^2$-s of solar flux hitting the surface, we'd need to cover ~89% of the area of the US (~10% of Titan's surface area) in photovoltaic cells, 8 × 10$^{12}$ m$^2$. (As an aside, photovoltaic cells covering 1.9 × 10$^{10}$ m$^2$, ~9% of the area of Kansas, would meet the energy needs of the US.) For comparison, the Solar Star photovoltaic power station in California has arrays covering 1.3 × 10$^7$ m$^2$.

Looking at this in another way, arrays covering a total of 8e12 m$^2$ on Titan would produce ~1.36 × 10$^6$ MW of power. This does not consider the efficiency with which solar energy is converted to useful energy (e.g. in a building). On Earth, such an array of cells would produce ~400X more energy; alternatively, 1.36 × 10$^6$ MW of power on Earth could be produced by a solar array ~1.94 × 10$^{10}$ m$^2$ in size.

One use of solar power could be to generate acetylene ($C_2H_2$) from methane ($CH_4$), using pyrolysis. Temperatures of ~1000°C would be required (Table 1 reaction 7). At higher temperatures (Table 1 reaction 7) $O_2$ could be generated by heating $CO_2$.

Future colonists on Titan employing solar cells for energy will likely need to be aware of seasonal rainfall, transient cloud cover (both of which are likely taken into account in our calculations that consider sunlight for 1/3 of the Titan day) and/or tholin sediments settling onto the panels.

### Geothermal

On Earth, remnant heat from planet formation and radiogenic heating is transported from the interior to the surface and can be accessed and utilized via geothermal plants. Significant heat has been measured at the surfaces of some solar system moons, including Enceladus, where Cassini instruments have observed at least ~4.7 GW of heat from fractures at the south pole [26] that is largely (though likely not completely) due to dissipation of tidal energy. Similarly, at Jupiter's moon Io, its heat flux of 2.25 W/m$^2$ is the result of intense tidal heating due to orbital resonances with Jupiter and other moons. At Titan, however, heat flux at the surface is not prominent.

The current outgassing of argon (40 Ar) as measured by Cassini INMS and GCMS is the best evidence for current-day outgassing on Titan [6]; such activity may also be responsible for the present-day atmospheric $CH_4$. There are features interpreted as hotspot volcanoes in the North Polar Region [27], but so far prominent hotspots have not been detected in Cassini data [28]. Titan's heat flux is just 5 mW/m$^2$ [29].

Presumably, if future missions to Titan are able to locate hot spot regions, these could potentially be used as sites to access geothermal energy; for now, since so little is known, and so little heat appears to be emanating from Titan, we assess the geothermal energy option as only a possibility in the future.

### Conclusion

In summary, Titan's natural resources present several options for useful energy sources for future visitors (or colonists/settlers) to the moon. Combustion of hydrocarbons alone is not efficient (due to the need to produce $O_2$ via electrolysis or other means), but hydrogenation of acetylene produces 376 kJ/mol, and hydrogenation of the abundant atmospheric nitrogen produces 92 kJ/mol. Other energy-producing chemical pathways are likely available.

An alternative to chemical energy is hydropower, utilizing Titan's large methane-ethane-filled seas and lakes to produce power as with hydropower here on Earth. Due to Titan's lower gravity, the power production is less than 20% of that on Earth, but the vast size of Kraken Mare, Titan's largest sea, would allow for power production for tens of thousands of Titan years. Detailed study of such an option requires improved knowledge of the topographic variations in and around the high latitude lakes and seas.

Wind power is a viable option. Wind speeds are low on Titan compared to Earth, but the atmospheric density is higher. Energy production would be relatively low at the surface of Titan, however the higher wind speeds at tens of km altitude could be accessed, e.g. by balloon-borne wind stations.

Solar power, due to the distance of Titan from the Sun and the atmospheric absorption, would be less efficient than on Earth, meaning that the arrays of solar panels would need to be larger by a factor of ~400. As on Earth, solar power on Titan is incredibly productive; we imagine that arrays of panels could be situated across the globe of Titan for ready access to power at all times of day and year.

Each of these sources of energy will be low compared to Earth, but if future humans on Titan use a combination of some or all of them, they'd have sufficient energy for heating and to derive breathable oxygen. Other options, such as tidal power and thermonuclear fusion, are possible but not discussed in this report.

Thus, while temperature, gravity and wind conditions on Titan mean that energy production efficiencies are generally lower than on Earth, Titan's wealth of natural resources nevertheless imply that energy production on Titan is a significant possibility.

## Acknowledgement

The authors are grateful to Ralph Lorenz and Jason Barnes for helpful comments on an earlier version of this paper, and to Aimee Oz for help with chemical energy calculations.